# *Weak measurement criteria for the past of a quantum particle*

B. E. Y. Svensson

*Theoretical High Energy Physics, Department of Astronomy and Theoretical Physics, Lund University, Sölvegatan 14, SE-22362 Lund, Sweden*

E-mail: Bengt_E_Y.Svensson@thep.lu.se

Vaidman (Phys.Rev. A **87**, 052104 (2013)) has proposed a "weak value" criterion for the past of a quantum particle, and applied it to photons in a particular setup of nested Mach-Zehnder interferometers. From his analysis, he draws some astonishing conclusions regarding which arms of the setup the photons populate. I argue that the weak values Vaidman employs as "weak tracers" in his analysis cannot be used in this case and that, consequently, the conclusions he draws cannot be upheld. The reason is that weak values are defined as a limit which in this case turns out to be discontinuous. I propose instead a "weak mean value" criterion which avoids these shortcomings.

## I. INTRODUCTION: VAIDMAN'S CRITERION

Vaidman – in [1], further elaborated on in [2-4] – has proposed a criterion for the past of a quantum particle in terms of "the weak trace it leaves". His aim is to improve the approach by Wheeler [5] which he [1] describes as follows: "It asserts that while we cannot discuss the past of a particle until it is measured, we can do so after the measurement. If the preselection led to a superposition of a few states and one of them was found in the postselection measurement, then we should regard the particle as being in the postselected state even before the postselection. Thus, the past of a quantum particle comes into existence due to measurement at a later time."

Wheeler's criterion is a non-local one; it involves more than the particular particle location one is interested in. One of Vaiman's aims is to find a more local criterion. An approach by means of a direct (projective) measurement at the location of interest, which would be the classical way of doing it, will of course not be possible in the quantum case, since it will destroy quantum coherence. But, and this is the *raison d'être* for Vaidman's proposal, a weak measurement can *both* be made locally *and* cause little disturbance to coherence, the less the weaker the measurement is.[1]

Although it is not stated explicitly, it is clear from Vaidman's writings that he means the "weak trace" to be understood operationally as the weak value of the projection operator onto that part of the set-up where one wants to establish the presence of the particle: A non-zero value for such a weak value would then, according to Vaidman, be a sign that the particle has been present at that location.

As an application of his criterion, Vaidman [1] treats the case of photons passing through a setup consisting of two nested Mach-Zehnder interferometers (MZI) as illustrated in Fig.1. For this setup, he arrives at some counterintuitive results. For example, he finds that there is a photon weak trace in the *B*- and *C*-arms without there being such a trace in either the *D*- or the *E*-arms: the photons seem to be able to appear in the inner MZI without entering or leaving it!

In this paper, I argue that these conclusions regarding the past of the photons in the nested MZIs cannot be upheld. The reason is that the limiting procedure that defines a weak value (*c.f.* Appendix 1 below) in the setup studied is discontinuous: the weak value deduced by going to the limit of vanishing weak measurement does not faithfully represent the undisturbed setup with no measurement. Therefore, the weak values cannot be used to draw any conclusions on the past of the photon in this case.

The plan of the paper is as follows. I start in section 2 by analyzing in detail the nested MZI setup of Fig. 1 in order to pinpoint exactly how and where the discontinuous limit occurs. I then devote a section to comment on the possible relevance to my analysis of the so called two-state vector formulation (TSVF) [10-12] of quantum mechanics (QM). Next, I venture to suggest an alternative criterion for the past of a quantum particle, still using weak measurements but avoiding the discontinuity that plagues Vaidman's weak value criterion. A short section gives an outline of a possible experimental realization of the new proposal. In the final section I give a summary of and some conclusions from my arguments. The appendices outline an operational definition of the weak value as well as my conventions for the unitary transitions in the beamsplitters of an MZI.

## II. ANALYSIS OF THE PHOTON STATE IN THE NESTED MACH-ZEHNDER INTERFEROMETERS

Referring to Fig.1, I denote the photon state in arm *A* by $|A>$, *etc.* for photon states in the other arms, and by $|D_1>$ through $|D_3>$ for the detector arms photonic states. The transition in the beamsplitter *BS1* is denoted *U(1)*, *etc.* for the other beamsplitters; the explicit expressions for these transitions may be deduced from eq. (B1) of Appendix B below. The inner interferometer in the undisturbed (*i.e.* without any measurement involved) setup is tuned so that no photons from the inner MZI enter into arm *E*. This is due to destructive interference

---

[1] For recent reviews of weak measurement and weak values see [6-9] with further references therein. I also include in Appendix 1 below a description of the operational definition of a weak value.

between the *B*- and *C*-arm photonic states in *BS3*. The photons in the innermost MZI thus all end up in the detector $D_3$.

The weak measurement – *c.f.* Appendix 1 – is described by a von Neumann measuring scheme [13] with a meter measuring the projection operator onto one of the arms of the MZIs. To be specific, let $\Pi_A = |A><A|$ denote the projection operator onto arm *A*, *etc.* for the other arms. Denote by $P_M$ the meter momentum observable conjugate to the meter pointer variable $Q_M$. The measuring device induces a unitary transition in the joint Hilbert space of the photon and meter which reads (the symbol $\otimes$ stands for the direct product of the photon and meter Hilbert spaces)

$$U_M(A) = \exp(-i\, g\, \Pi_A \otimes P_M) = 1 - \Pi_A \otimes [\exp(-i\, g\, P_M) - 1]$$

$$\approx 1 - i\, g\, \Pi_A \otimes P_M. \qquad (1)$$

Here, the last approximation is valid in the weak limit when the measurement coupling strength *g* is small. Expressions similar to (1) hold for the other arms.

Following Vaidman [1], the weak trace is now operationalized to mean a nonvanishing weak value of the projection operator onto that arm of the setup one is interested in.

A weak value requires specification of an initial, "preselected" state $|in>$ and a final, "postselected" state $|f>$. In the example to be treated, $|in>$ is the state $|N>$ (corresponding physically to a photon entering the MZI setup from the photon source), while $|f>$ is taken to be $|D_2>$ (corresponding physically to a click in the detector $D_2$).

Let me now deduce the weak value for the projection operator onto the *B*-arm. On Vaidman's criterion, this weak value being nonzero means that the photon has been present in arm *B* of the nested MZIs. The calculation I present uses a standard QM approach; in the next section I comment on any possible bearing the two-state vector formalism [10-12] might have on the analysis.

I present this standard analysis by exhibiting how the joint photon-meter initial state $|in> \otimes |m>$ (with $|m>$ the initial state for the meter) evolves forward in time through the setup, also including a (weak) measurement of the projector onto the *B*-arm. This evolution reads

$$|in> \otimes |m> \rightarrow U(1)|in> \otimes |m> \tag{2a}$$

$$\rightarrow U(2) U(1) |in> \otimes |m> \tag{2b}$$

$$\rightarrow U_M(B) U(2) U(1) |in> \otimes |m>$$

$$= U(2) U(1) |in> \otimes |m>$$

$$+ [U_M(B) - 1] U(2) U(1) |in> \otimes |m> \tag{2c}$$

$$\rightarrow U(3) U(2) U(1) |in> \otimes |m>$$

$$+ U(3) [U_M(B) - 1] U(2) U(1) |in> \otimes |m> \tag{2d}$$

$$\rightarrow U(4) U(3) U(2) U(1) |in> \otimes |m>$$

$$+ U(4) U(3) [U_M(B) - 1] U(2) U(1) |in> \otimes |m>. \tag{2e}$$

More explicitly, and using the conventions for the beamsplitter transitions $U(1)$ etc as exhibited in Appendix 2, an expression like $U(2) U(1)|in>$ stands for

$$U(2) U(1) |in> = -i (\sqrt{2}|A> + |B> + i|C>)/2, \tag{3}$$

and $U(3) U(2) U(1) |in>$ for

$$U(3) U(2) U(1) |in> = -i (|A> + |D_3>)/\sqrt{2}. \tag{4}$$

In fact, what matters for the discontinuity argument that I am about to present is only the main structure of these expressions, *i.e.*, which photon arm states do occur, not their respective coefficients. Of course, these coefficients are important for calculating numerical expressions for the weak values.

The two terms in lines (2c -e) have different contents. Consider the line (2d). The first term there represents the undisturbed case with, from (4), photonic states in the *A*- and $D_3$- arms only. In particular, destructive interference in *BS3* between the *B*- and *C*-arm states means no photon state into the *E*-arm. The second term represents the influence of the measurement and does have an amplitude into the *E*-arm: the factor $U_M(B) - 1$ is, according to (1), proportional to $\Pi_B$, implying that this second term, representing a state in the *B*-arm only, is not subject to the destructive interference with the *C*-arm state in *BS3*.

To get the meter to reveal the weak value, one has to perform a (projective) measurement on the photons, projecting the state (2e) onto the postselected state $|f> = |D_2>$. The (un-normalized) meter state after this post-selection is then

$$< f \,|\, [\ U(4)\ U(3)\ U(2)\ U(1) |\ in > \otimes\ |\ m >$$

$$+\ U(4)\ U(3)\ \Pi_B\ U(2)\ U(1) |\ in > \otimes\ [\exp(-i\,g\,P_M) - 1] \,|\, m >]$$

$$\approx\, < f\,|\, U(4)\ U(3)\ U(2)\ U(1) |\ in >\ [\ 1 - i\,g\ {}_f\{\Pi_B\}_w\ P_M\ ] \,|\, m >, \qquad (5)$$

where the last line is the weak approximation and where the weak value ${}_f\{\Pi_B\}_w$ of the projection operator onto the B-arm is given by

$${}_f\{\Pi_B\}_w\ =\ <f\,|\,U(4)\ U(3)\ \Pi_B\ U(2) U(1) |in> \,/\, <f\,|\,U(4)\ U(3)\ U(2)\ U(1)\ |\ in>. \qquad (6)$$

According to Vaidman [1], the non-vanishing of this weak value is interpreted as the presence of a photon in the B-arm.

I note in passing that it is the fact that you have a non-zero state in the A-arm that allows you to extract a weak value for the number/projection operator in the B-arm: blocking the A-arm and you cannot extract such a weak value. This shows the quantum mechanical interconnectedness, the "non-locality", of the setup. I comment further on the locality issue in section IV below.

It is clear from this detailed calculation that *it is the very act of performing a weak measurement + postselection that gives a nonzero result for this weak value: the measurement, even if ever so weak, "disturbs" the photon state* by allowing part of the B-arm photon state to pass un-interfered through the BS3 beamsplitter into the E-arm. This then ultimately results in a signal in the $D_2$-detector implying a nonvanishing weak value for the B-arm projection operator.

Another way of looking at this result is to observe (see Appendix 1) that the weak value ${}_f\{\Pi_B\}_w$ is obtained by a limiting procedure involving the mean value after postselection, ${}_f<Q_M>$, of the meter pointer variable $Q_M$:

$${}_f\{\Pi_B\}_w\ =\ \lim_{g \to 0}\ [\ (1/g)\ {}_f<Q_M>\ ]. \qquad (7)$$

This limit is to be taken using the expression (2e), all the time with $g \neq 0$, i.e., all the time with a photonic state in the E-arm arising from the measurement of the B-arm projector. For small g, the amplitude of this E-arm state is indeed tiny, being linear in g. But the fact that you divide by g to get the weak value means that you may in any case get a non-zero weak value however tiny this amplitude. But the situation for $g = 0$ is different. Then there is no second term in the expression (2e), so no state corresponding to a photon in the E-arm. The limit $g \to 0$ is thus a discontinuous one in the sense that the limit is taken with a contribution from an E-arm state while there is no such contribution when $g = 0$, i.e., in the situation undisturbed by any measurement. *The limiting procedure defining the weak value is discontinuous when you compare the configuration used in taking the limit $g \to 0$ to the undisturbed configuration $g = 0$.* [2]

---

[2] An analogy to this situation, although not a perfect one, is an ordinary function $f(x)$ which for small $x > 0$ has an expansion $f(x) = a\,x + O(x^2)$ but which for $x = 0$ has a non-vanishing value. Then $\lim_{x \to 0} [(1/x)\,f(x)] = a$, while for $x = 0$ the expression $\frac{1}{x} f(x)$ is undefined (or, if you prefer, infinite).

The conclusion is clear. You are not allowed to conclude from the case *g > 0* to the *g = 0* unless the limit *g → 0* is continuous. Since this is not the case here, the weak value cannot represent the photon behavior in the innermost MZI undisturbed by measurement. *Vaidman's "weak trace" criterion cannot be applied to this case due to the discontinuous behavior of the limiting procedure in getting the weak value.*

It goes without saying that Vaidman's far-reaching arguments against "the common sense description" of the past of a quantum particle, based as they are on the weak values, lose their very foundation. The discontinuous behavior invalidates his arguments. Statements (quoted from [4]) like "The photons tell us that they have been in the parts of the interferometer through which they could not pass." and "… (the photons) never left the nested interferometer…" become unfounded: they rely on the weak value as the relevant entity in the weak trace criterion. As I have shown, however, the weak values do not give a faithful picture of the (undisturbed) MZI setup.

### III. THE TWO-STATE VECTOR FORMULATION APPROACH

In his articles, Vaidman makes heavy use of the so called two-state vector formalism TSVF [10-12]. In this formalism, a quantum state between two measurements is described by a "two-state" vector $<\Phi\,||\,\Psi>$. Here, the second of the two states, $|\Psi>$, is the usual quantum state evolving forward in time from the initial, pre-selected state. As an example, in the expression (6) it is given by $|\Psi> = U(2)U(1)|in>$. The first state $<\Phi|$ is the Hermitian conjugate (denoted †) of the quantum state evolving backwards in time from the final, post-selected state. In the expression (6), it is given by $<\Phi| = [U(3)^{-1} U(4)^{-1}|f>]^{\dagger}$. This formalism is strongly advocated by Vaidman as an appropriate way of understanding, visualizing and talking about the weak trace of his criterion.

In fact, the TSVF description of the nested MZI setup is tailored to a description in terms of weak values of projection operators. A statement in TSVF can be directly translated into a statement of a weak value of such an operator and conversely. From a logical point of view, then, TSVF is equivalent to an analysis in terms of weak values. Therefore, the TSVF approach is subject to the same arguments as I have put forward in the previous section.

Nor can the experimental results of [4] (see also Section V below) have any bearing on the arguments I have presented. An experiment, however interesting *per se*, can never have anything to say about the *interpretation* of its results. Only a theory can.

## IV. AN ALTERNATIVE WEAK TRACE CRITERION FOR THE PAST OF A QUANTUM PARTICLE.

The idea of formulating a "weak trace" criterion for the past of a quantum particle is, however, too good to be given up that easily. In fact, I think there is another way of invoking weak measurement for this purpose: use *weak measurement but without postselection* and let the "weak trace" be identified with the mean value of the projector onto the location of interest. Operationally, in the (weak) measurement scheme, this mean value is essentially the mean value of the meter observable, eq. (11) below. I call this "the weak mean value" criterion.

Indeed, one of Vaidman's intentions in proposing the weak value for the weak trace criterion is to find a more local criterion for the past of a QM particle than, *e.g.*, Wheeler's [5]. But on this point he is not completely successful. Since a weak value depends also on the postselected state, the weak value criterion obviously involves more than what occurs at the location of interest. In other words, the weak value criterion is not as local as one might have wished. If one instead, as I propose, identifies the weak trace with a non-vanishing weak *non-postselected mean value* of a projection operator, the criterion will be as local as is possible.

For the nested MZI setup discussed previously, an application of this "weak mean value" criterion goes as follows.

Start with the line (2c) above and rearrange it slightly. Use

$$[\Pi_A + \Pi_B + \Pi_C] = 1 \tag{8}$$

to write

$$U_M(B) = [\Pi_A + \Pi_C] + \Pi_B \otimes \exp(-i g P_M). \tag{9}$$

Then one finds

$$|in> \otimes |m> \rightarrow U_M(B) U(2) U(1) |in> \otimes |m>$$

$$= [\Pi_A + \Pi_C] U(2) U(1) |in> \otimes |m>$$

$$+ \Pi_B U(2) U(1) |in> \otimes \exp(-i g P_M) |m> \tag{10}$$

for the state which, in the forward time evolution, describe the photon before the innermost MZI states enter the beamsplitter *BS3*.

Next, calculate the mean value $<Q_M>$ of the meter variable in this state (assuming, as I do, that this mean value is zero in the initial meter state). One finds

$$< Q_M > = <m | \exp( + i g P_M ) Q_M \exp( - i g P_M )| m >$$

$$\times < in | U(1)^{-1} U(2)^{-1} \Pi_B U(2) U(1)| in > =$$

$$= g < in | U(1)^{-1} U(2)^{-1} \Pi_B U(2) U(1)| in >. \tag{11}$$

Note that this result is exact: no small-$g$ approximation is invoked. In fact, the von Neumann measurement of the projection operator affects only the non-diagonal terms (those proportional to $|A><B|$, *etc.*) of the photon density matrix.

The conclusion is that the (non-postselected) mean value $< Q_M >$ of the meter observable is a direct measure of the presence of the photon in the *B*-arm as manifested by the second factor in (11).

The non-vanishing of the expression

$$lim_{g \to 0} [ (1/g) < Q_M > ] = < in | U(1)^{-1} U(2)^{-1} \Pi_B U(2) U(1)| in > \tag{12}$$

may then be taken as a weak trace criterion for the presence of photons in the *B*-arm. In particular, even if (11) is exact, a weak measurement, *i.e.*, a small-$g$ approximation, is needed in order for the measurement not to disturb (more than minimally) the quantum coherence of the succeeding time evolution.

One should note the local character of this criterion: as it stands, it is independent of any perturbation that the measurement may induce to the succeeding time-evolution of the photon state. So from the point of view of merely measuring the mean value of the projection operator $\Pi_B$, it does not matter whether, for example, there is a transition into the *E*-arm from *BS3*. The meter attached to the *B*-arm to measure the mean value of $\Pi_B$ has so to speak already registered what it is supposed to register and need not wait for a post-selection to occur in detector $D_2$. Of course, unless it is weak, the $\Pi_B$-measurement will disturb that succeeding time evolution, so if one wants to keep intact as much of the original evolution as possible, the measurement has nevertheless to be weak.

This more strict locality property also implies that there is no discontinuity involved in taking the limit (12) as compared to the undisturbed, no-measurement situation.

If the "weak trace" is interpreted with this "weak mean value" criterion, one may also convince oneself that the "common sense' interpretation of the past of a photon in the nested MZI setup is restored. There is no question of "the photon (leaving) a trace in a path through which it did not pass" as stated in [1], nor for the photons to "have been inside the nested interferometer .. (which) they never entered and never left…" as it was stated in [4].

In sum, a weak trace scheme with a weak mean value criterion has definite advantages. Since it involves weak measurements, its influence on the undisturbed setup is minimal. Since it does not involve any postselection, it has the property of being as local as possible. And it introduces no discontinuity. It seems therefore to be an ideal way of establishing, through (weak) measurements, the trace of a QM particle.

## V. EXPERIMENTAL REALIZATION OF THE WEAK MEAN VALUE CRITERION.

In [4], Danan et al present an interesting experimental realization of the nested MZI setup suggested by Vaidman. Their setup is slightly different from that of Fig.1 – see [4] for details – but the main characteristics remain the same. In particular, it would correspond to using vibrating mirrors *M1* – *M3* to implement (weak) measurements. The meter pointer variable is then the deviation of the (transverse) momentum of the photon beam from the fixed-mirror situation. For example, a vibrating mirror *M1* with the mirrors *M2* and *M3* fixed will constitute a measurement of the presence of photons in the *A*-arm. In fact, Danan *et al* let the frequency of the vibrations be different for the different mirrors in order to identify photons in the respective arms of the setup. To define the postselected state, they place a (differential) detector in the position $D_2$ of Fig.1. A signal at the relevant frequency of that detector is then a measure of the weak value of the projection operator onto the arm of the respective vibrating mirror. – See [4] for all the details.

This general procedure of Danan *et al* can rather straightforwardly be adapted to provide an experimental realization also for the suggested "weak mean value" criterion for the past of a QM particle. The principle behind such an adaption is as follows.

Suppose you vibrate the mirrors *M1* through *M3* like Danan *et al* do. Also, let the meter pointer observable be the (transverse) momentum deviation, called *y* by Danan *et al*. Then, replace the differential detector $D_2$ in the Danan *et al* experiment with (non-differential) detectors at the positions $D_1$ through $D_3$ of Fig. 1. These detectors shall directly register the deviation *y*. Taking as an example the tracing of the photon in the *B*-arm, you will have to vibrate the mirror *M2* and read off the *y*-value in detector $D_3$. The detector should, in particular, deliver the mean value of this deviation, as a direct measure – *c.f.* eq. (11) – of the weak mean value of the projector $\Pi_B$.

## VI. SUMMARY AND CONCLUSIONS

This paper is devoted to some aspects of identifying the past of a quantum particle. In particular, I have focused on Vaidman's "weak trace" criterion [1] in which the past of a quantum particle is revealed by a nonzero weak value of the projection operator onto that location which one is interested in. For the particular example studied – photons in the nested Mach-Zehnder setup of Fig.1 – I show that the weak values cannot be used to characterized the setup when it is undisturbed by the measurement. This is due to the fact that even an ever so weak measurement disturbs the photonic state in a discontinuous way. The limiting procedure with weaker and weaker measurement strength is, in fact, discontinuous with respect to the situation with no measurement. As a consequence hereof, and of the fact that a weak value is defined by this limiting procedure – see eq. (7) – the weak values do not any longer characterize the (undisturbed) setup and cannot be used to draw any conclusions on the past of the photons for this setup.

A treatment using the two-state vector formalism TSVF cannot come to the rescue. This is so because the TSVF is but a reformulation of a weak value one. Logically, the two formulations are equivalent.

It follows that the rather far-reaching conclusions that Vaidman and his collaborators draw regarding "photons (that) leaves a trace in a path through which they did not pass" (quoted from [1]) cannot be upheld. Instead of inferring that "the past of the photons is not represented by continuous trajectories" ([4]), one must rather note that it is the very definition of the weak value as a limit that introduces a discontinuity.

As an alternative to Vaidman's "weak value" criterion I propose a "weak mean value" criterion which does not run into the problems of Vaidman's criterion. It does so by avoiding the dependence on a post-selected state which a weak value has. The proposed criterion does not lead to any of the "non-commonsense" traits found by Vaidman. From the point of view of experimental realization, the "weak mean value" criterion can be examined by a slight variation of the apparatus used by Danan *et al* [4] to examine Vaidman's "weak value" criterion.

## ACKNOWLEDGEMENT

I am indebted to Ruth Kastner for drawing my attention to the paper by Danan *et al* [4], leading to my interest in the articles [1-3], and for valuable comments.

## APPENDIX A: OPERATIONAL DEFINITION OF A WEAK VALUE.

I here sketch the main steps in the experimental procedure to arrive at a weak value. A more thorough treatment – with the necessary caveats – can be found, *e.g.*, in [6]; see also [7-9] or any of the relevant references cited therein.

(i)   Let the system under study, $\mathcal{S}$, be subject to a weak von Neumann type measurement [13] of one of its observables, $S$, by a measuring device, a 'meter' $\mathcal{M}$. Think of the procedure as being done on one representative of $\mathcal{S}$ after the other, be it an ordinary particle, a photon or whatever physical system is considered. – In the example studied in the main text above, the system $\mathcal{S}$ is a photon and the observable $S$ is a projection operator onto one of the arms of the MZIs.

(ii)  The initial system-meter state is assumed to be $|in> \otimes |m>$. Here, $|in>$ is the initial, "preselected" state for the system and $|m>$ is the initial state of the meter, while the symbol $\otimes$ denotes the direct product of the system and the meter Hilbert spaces. The meter initial state is chosen so that its wave function $<q|m>$ – with $|q>$ an eigenstate of the meter pointer observable $Q_M$ – is a Gaussian peaking at $q=0$ and width (in its square) of $\Delta$. In particular, the mean value $<Q_M> \equiv <m|Q_M|m>$ of $Q_M$ in the initial state is assumed to vanish.

(iii)  Invoking a von Neumann measurement scheme [13], a measurement interaction is described as a transformation of this state into the state $U[|in> \otimes |m>]$ with $U = \exp(-i\,g\,S \otimes P_M)$, where $P_M$ is the momentum conjugate to the meter pointer observable $Q_M$. A weak measurement is characterized by a small value of the measurement strength $g$, allowing an expansion in $g$ so that $U = 1 - i\,g\,S \otimes P_M + O(g^2)$. – Vaidman, in [1] and earlier papers, puts $g = 1$, and instead uses (an entity proportional to) $1/\Delta$ as the small expansion parameter. However, I find it convenient to keep $g$ as the regulating parameter to be able to talk easily of the strength of the interaction and to let $g = 0$ characterize the non-measurement situation. The two approaches are equivalent as long as $g \neq 0$.

(iv)  Next, project the system state onto a "postselected", final state $|f>$.

(v)  Finally, submit the measuring device to a projective measurement, resulting in a definite value $q$ of the pointer variable.

(vi)  Repeat to get enough statistics.

(vii)  The ensuing mean value $_f<Q_M>$ of the pointer observable, representing the shift in the peak of the meter Gaussian wave function with respect to its initial position, will then, for small values of $g$, be given by $g\,_fS_w$ with the weak value $_fS_w$ of $S$ defined by

$$_fS_w = <f|S|in>/<f|in>, \qquad (A1)$$

which means that

$$_fS_w = \lim_{g \to 0} [(1/g) \,_f<Q_M>]. \qquad (A2)$$

Here, I prefer the slightly clumsy notation with a pre-subscript $f$ attached to the relevant symbols in order to emphasize that the respective entities require specification also of the post-selected state $|f>$.

Two remarks to this scheme are appropriate.

The first is that a single measurement does not provide much information about the system. But after a large number of repeated measurement – point (vi) above – on an ensemble of identically prepared pre- and postselected systems, useful information can be extracted.

The second is that the weak value $_fS_w$ is defined by a limiting procedure (A2). In most cases, this limit is continuous and the limiting procedure causes no problems. However, in the case with the nested Mach-Zehnder interferometers studied in the main text above, this limit is discontinuous and requires special considerations.

**APPENDIX B: CONVENTION FOR TRANSITION IN A BEAMSPLITTER.**

In the notation of Fig.1, and for the beamsplitter *BS3* in front of the detector *D₃* – assumed as all the other beamsplitters to be well-balanced, 50-50 ones – my convention amounts to the unitary transition

$$\begin{pmatrix} |D_3> \\ |E> \end{pmatrix} = U(3) \begin{pmatrix} |B> \\ |C> \end{pmatrix} = 1/\sqrt{2} \begin{pmatrix} 1 & i \\ i & 1 \end{pmatrix} \begin{pmatrix} |B> \\ |C> \end{pmatrix}, \qquad (B1)$$

with similar expressions for the other beamsplitters. Any other consistent convention for the transition matrix will give equivalent results.

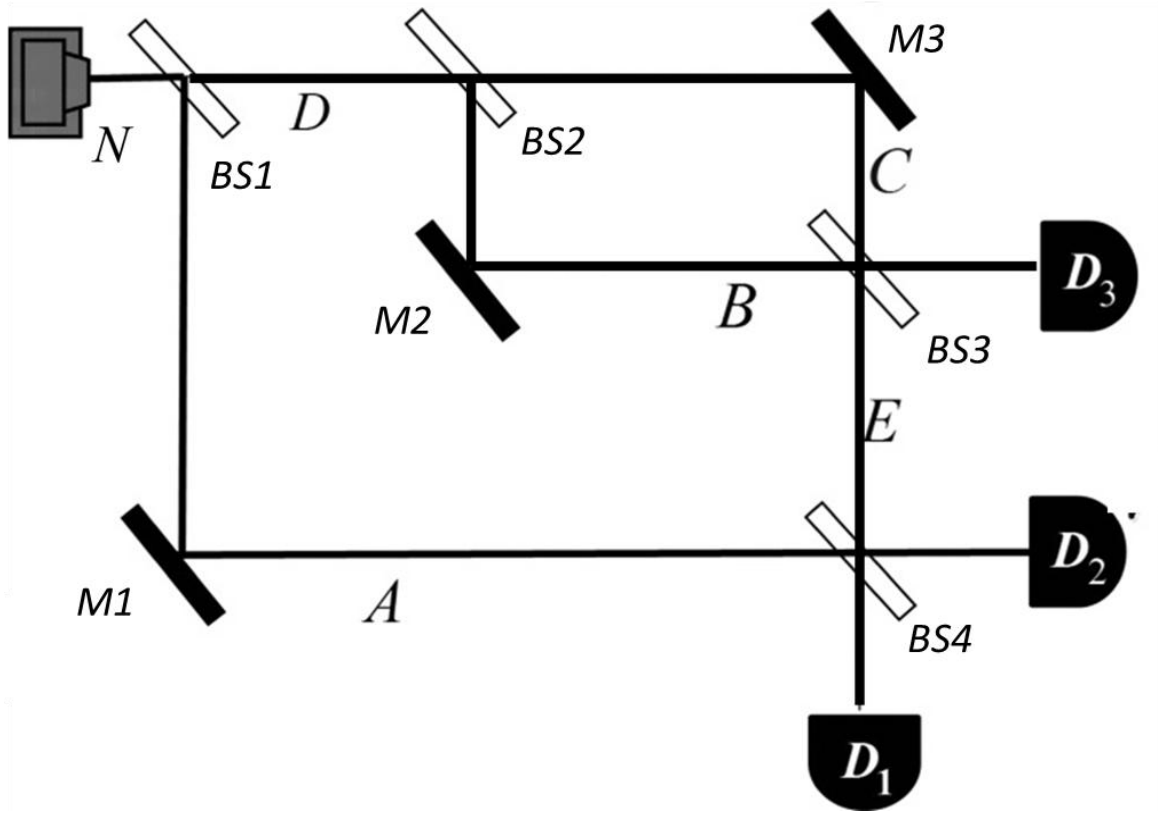

FIG.1. The nested Mach-Zehnder interferometer setup. The arm directing the photons from the photon source into the set-up is denoted N. Symbols *A* through *E* denote the interferometer arms, *BS1* to *BS4* are beamsplitters, *M1* to *M3* are mirrors, and $D_1$ to $D_3$ are detectors.(Adapted from Vaidman [1])